\documentclass[aip,amsmath,amssymb,reprint,floatfix]{revtex4-1}

\usepackage{graphicx}% Include figure files
\usepackage{dcolumn}% Align table columns on decimal point
\usepackage{bm}% bold math

\newcommand{\YbFour}{^{174}\text{Yb}^+}

\begin{document}
\title{An Integrated Mirror and Surface Ion Trap with a Tunable Trap Location}
\author{Andre Van Rynbach}
\email{ajv6@duke.edu}
\affiliation{Department of Electrical and Computer Engineering, Duke University, Durham, North Carolina 27708, USA}
\author{Peter Maunz}
\affiliation{Sandia National Laboratories, P.O. Box 5800, Albuquerque, New Mexico 87185, USA}
\author{Jungsang Kim}
\affiliation{Department of Electrical and Computer Engineering, Duke University, Durham, North Carolina 27708, USA}
\date{\today}

\begin{abstract}
We report a demonstration of a surface ion trap fabricated directly on a highly reflective mirror surface, which includes a secondary set of radio frequency (RF) electrodes allowing for translation of the quadrupole RF null location.  We introduce a position-dependent photon scattering rate for a $^{174}$Yb$^+$ ion in the direction perpendicular to the trap surface using a standing wave of retroreflected light off the mirror surface directly below the trap.  Using this setup, we demonstrate the capability of fine-tuning the RF trap location with nanometer scale precision and characterize the charging effects of the dielectric mirror surface upon exposure to ultra-violet light.
\end{abstract}

% \pacs{some PACS}%
%\keywords{}

\maketitle

An important property of trapped ion qubits for the scalability of quantum computing systems is their capability for use in conjunction with photonic interconnects.  Collecting entangled photons spontaneously emitted from an ion into a single mode fiber provides a means to entangle a pair of atomic ion qubits in two remote locations and create distributed entangled states for a quantum network. \cite{Moehring:2007aa}  By distributing entangled qubit pairs over several individual quantum processing units, one can potentially overcome a limitation on the total number of trapped ions that can be combined into a single processing unit. \cite{PhysRevA.89.022317}  Effective integration of an optical cavity with an ion trap is a long standing goal in the field towards realizing dramatic increases in the efficiency and fidelity of these photonic interconnects. \cite{single_photons,Sterk,Stute:2012aa,Yb-cavityQED}   Fabricating a surface ion trap directly on a highly reflective (HR) mirror allows for the ion trap to be combined with a small mode volume optical cavity to create an efficient photon collection platform for improving entanglement generation rates between remote ions\cite{Herskind:11, taehyunCavity} and achieving faster quantum state detection.\cite{Horak_detect}  Here we present results on the fabrication and testing of an integrated mirror and surface ion trap and demonstrate the ability to fine tune the trap location of the ion perpendicular to the trap surface with nanometer scale precision.  This trap design also allows for fine tuning of the trap location parallel to the trap surface using additional control electronics.

\begin{figure}[b]
	\includegraphics[width=\columnwidth]{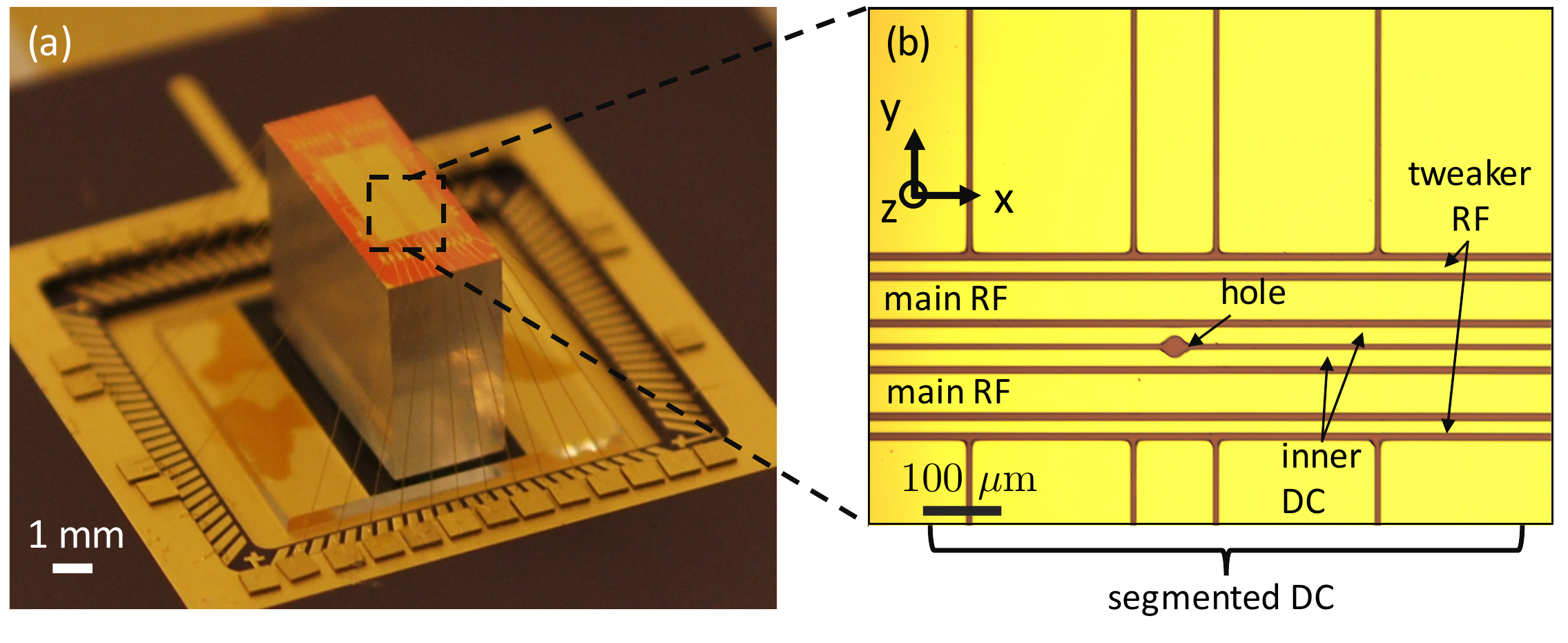}
	\caption{ (Color online)  (a) Image of the mirror trap mounted and wirebonded to the CPGA package.  (b) Close up image of the trap electrodes (gold colored regions) around the 24 $\mu$m hole for the cavity. }
	\label{fig:trap}
\end{figure}

Much progress has been made towards integrating optical components with surface traps, including mirrors,\cite{GATechMirror} lenses,\cite{PhysRevLett.106.010502}, optical fibers,\cite{PhysRevLett.105.023001,chuang_fiber_trap} and integrated waveguides. \cite{Chiaverini}  For optical cavity integration, the ability to precisely position the trap location at an antinode of the cavity mode is necessary to maximize the coupling strength between the ion and the cavity.  In standard surface traps, the ideal trapping location is characterized by a vanishing of the radio frequency (RF) field, called the RF null, and is determined by the geometry of the RF electrodes.  In a linear surface trap, the height of the trapping location can be estimated from the spacing and width of the linear RF electrodes.\cite{Tanaka2011}  To make the trap height an adjustable parameter, we have added a second set of RF electrodes, which we call the ``tweaker electrodes," to provide the capability of adjusting the ion height without moving away from the RF null and thereby avoiding the introduction of excess micromotion (Fig.~\ref{fig:trap}b). \cite{berkeland_micromotion}  Adding multiple sets of RF electrodes has been implemented in surface traps previously using an adjustable capacitive coupling to vary the ratio of voltages between the various RF electrodes to adjust the trap height with micron scale resolution. \cite{rf_null_move,PhysRevLett.105.023001}  Our approach using a separate RF source allows for continuous adjustment of the trap height with a resolution of several nanometers by tuning the relative voltage amplitude between the main RF and the tweaker RF.  In addition to integrated optics and cavities, this capability has broader applicability to standing wave laser gates\cite{GaTechSW} and multi-well traps.\cite{tanaka_multiwell}

We fabricated a surface ion trap directly on a HR ultra-violet (UV) mirror with the goal of using it in a hemispherical cavity arrangement with trapped ytterbium ions. \cite{taehyunCavity}  The rectangular fused silica mirror substrate has dimensions of 8 mm $\times$ 3.25 mm with a thickness of 6.35 mm.  The HR mirror coating was made from a dielectric stack (deposited by Advanced Thin Films) and optimized for 369.5 nm wavelength light.  We developed a photolithography and metal deposition process to pattern the trap electrodes directly on this mirror surface.  We used a high spin speed of 6000 rpm for spin coating the mirror with a negative photoresist (NFR016D2) to reduce the surface tension build up of photoresist on the edges of the substrate.  The photoresist was exposed at an intensity of 14 mW/cm$^2$ with 365 nm light for 7.2 seconds, followed by a post exposure bake at 90$^\text{o}$ C for 165 seconds and a development step with MF-319 developer for 60 seconds.  The roughly one third reduction in exposure time compared to the standard process with this photoresist is due to reflection of the exposure light from the mirror surface.  Following the photolithography step, an e-beam metal evaporation step was performed to deposit 20 nm of Ti followed by 350 nm of Au.  The metal liftoff was achieved by soaking the sample in acetone and rinsing with isopropanol.  Small capacitors of approximately $820$ pF were placed around the edge of the ceramic pin grid array (CPGA) package with Epo-Tech H20E conductive epoxy to filter out any RF pickup on the direct current (DC) electrodes (Fig.~\ref{fig:trap}a).

Our trap design includes a hole in the inner DC electrodes to expose the underlying mirror surface below the trap (Fig.~\ref{fig:trap}b).  We experimentally verified with a test cavity that a hole size of 24 $\mu$m was a good compromise between limiting the exposed area of the ion to a dielectric surface underneath the trap and minimizing clipping losses of the cavity mode.  By measuring the finesse of this cavity, we determined the photolithography fabrication process did not lead to a measurable decrease in the mirror coating quality when compared to a cavity formed with a bare mirror with the same coating.  The main RF electrodes in the mirror trap are 57 $\mu$m wide and separated by 60 $\mu$m, leading to a nominal ion trap height of 50 $\mu$m above the trap surface.  A $5$ $\mu$m gap between the RF electrodes and neighboring DC electrodes is used to prevent a leakage current. The main RF voltage is amplified using a resonant LC circuit made from a helical resonator with frequency around 40 MHz.  We drive the tweaker electrodes using a high bandwidth, high slew rate operational amplifier (op amp, ADA4870) with a sinusoidal voltage generated by a direct digital synthesizer (DDS).  The phase and frequency of the DDS output can be digitally controlled to match those of the main RF trap voltage, and its amplitude is set with a 10-bit number between 0 and 1,023.  For the tweaker electrode width of 20 $\mu$m and a main RF voltage amplitude of 185 V, our trap simulations predict the ion trap height increases linearly with the applied voltage with a slope of 37 nm/V when the phase and frequency are matched to those of the main RF signal.  If the two tweaker electrodes are driven by independent RF sources with different amplitudes, the trap location can be shifted along the y-axis as well.  A low pass filter is used after the op amp to filter out higher order harmonics of the amplified voltage signal as it drives the primarily capacitive load of the trap (capacitance $\sim12$ pF).  We measure the voltage applied to the tweaker electrodes with a capacitive voltage divider placed at the amplifier output in parallel with the trap with a pickoff ratio of 220:1.  This voltage divider is also used to calibrate the relative phase between the two RF signals, as the capacitive pickup of the main RF on the tweaker electrodes can be measured with this pickoff using a high impedance oscilloscope.  Any phase mismatch between the main RF and tweaker RF signals will introduce excess micromotion of the ion as the net electric field will no longer have a fixed RF null location.

\begin{figure}[b]
	\includegraphics[width=\columnwidth]{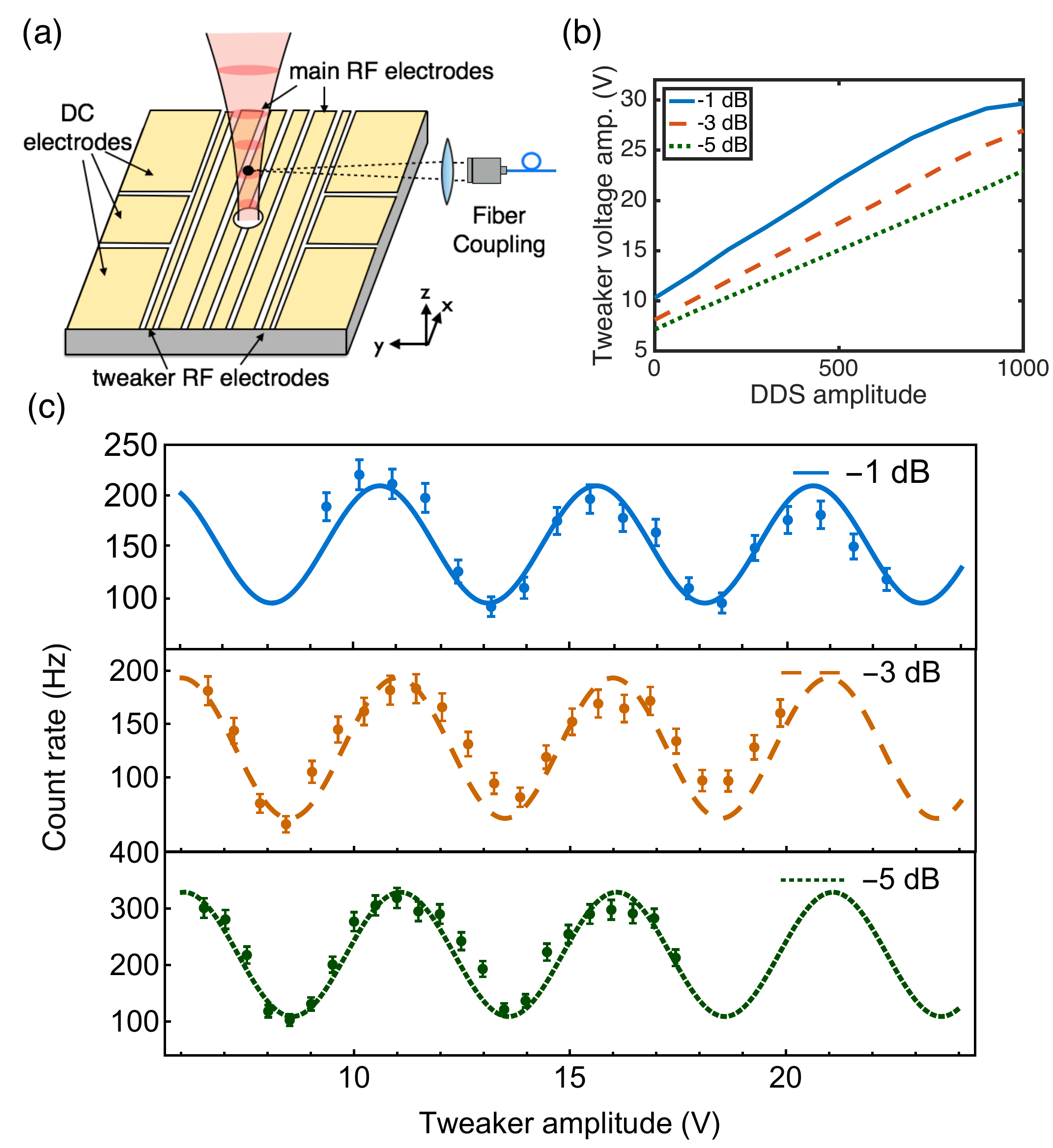}
	\caption{(Color online) (a) Schematic of the experimental setup showing the standing wave created by retroreflecting a laser beam off the trap surface and the photon collection optics from the side of the ion trap into a single mode fiber. (b) Voltage measurements of the tweaker amplitude for various levels of input power attenuation.  The amplifier output saturates above an amplitude of about 23 V. (c) Oscillations in the ion scattering rate as a function of the voltage on the tweaker electrodes for various levels of input attenuation for a SW laser detuning of $\Delta=-2\pi \times 10$ MHz.  We fit the data to a sine function to extract the periodicity of the oscillations which correspond to a translation of $\lambda/2\simeq 185$ nm.}
\label{fig:oscillations}
\end{figure}

We observed stable trapping of $\YbFour$ ions in the mirror trap for several hours at a time with Doppler cooling using the $^2\text{S}_{1/2}$$\leftrightarrow$$ ^2\text{P}_{1/2} $ transition at a wavelength of 369.5 nm and a repump beam at 935 nm wavelength for pumping out of the $^2 D_{3/2}$ state. \cite{Yb171}  We locate the ion position along the z-axis by setting up a standing wave (SW) from the retroreflection of 369.5 nm wavelength light off the mirror surface under the trap (Fig.~\ref{fig:oscillations}a).  This setup provides a spatially varying light intensity perpendicular to the trap surface with a periodicity of $\lambda/2 \simeq 185$ nm.  As the trap location is moved by driving the tweaker electrodes, the modulation of the scattering rate of the photons in the SW by the ion provides a quantitative scale for the ion's location perpendicular to the trap surface.  The scattered photons are collected from the side of the trap using a single mode fiber to spatially filter out background photons from laser light scattering off the trap structures and are detected with a photomultiplier tube (PMT, Hamamatsu H10682-110).  This allows us to reduce the background counts to only about 10 counts per second, which is limited by the dark count rate of the PMT.

A complete description of the ion's fluorescence in the SW must take into account the effects of the micromotion on the scattering rate.  If we transform to the ion's frame of reference, the SW laser will appear as if it is phase modulated at the RF frequency due to the presence of any micromotion along the z-axis.  This will give rise to frequency sidebands at integer multiples of the RF frequency, $\Omega=2\pi \times 42.5$ MHz.  We can subsequently write the SW laser's electric field at the ion location for a micromotion amplitude of $a_{z}$, laser frequency $\omega_0$, laser wavenumber $k$, and electric field amplitude $E_0$ as 
\begin{equation}
\label{eq:mod_field}
E(z,t) = 2 E_0 \sin(k z) \exp(-i \omega_0 t) \exp(i k a_{z} \sin(\Omega t)).
\end{equation}
The second exponential term can be expanded in terms of Bessel functions as 
\begin{equation}
\label{eqn:bessel}
 \exp(i \beta \sin(\Omega t)) = \sum_{n=-\infty}^{\infty} J_n(\beta) \exp(i n \Omega t),
\end{equation}
where we have defined the modulation parameter $\beta=k a_{z}$.

\begin{figure}[b]
	\includegraphics[width=0.90\columnwidth]{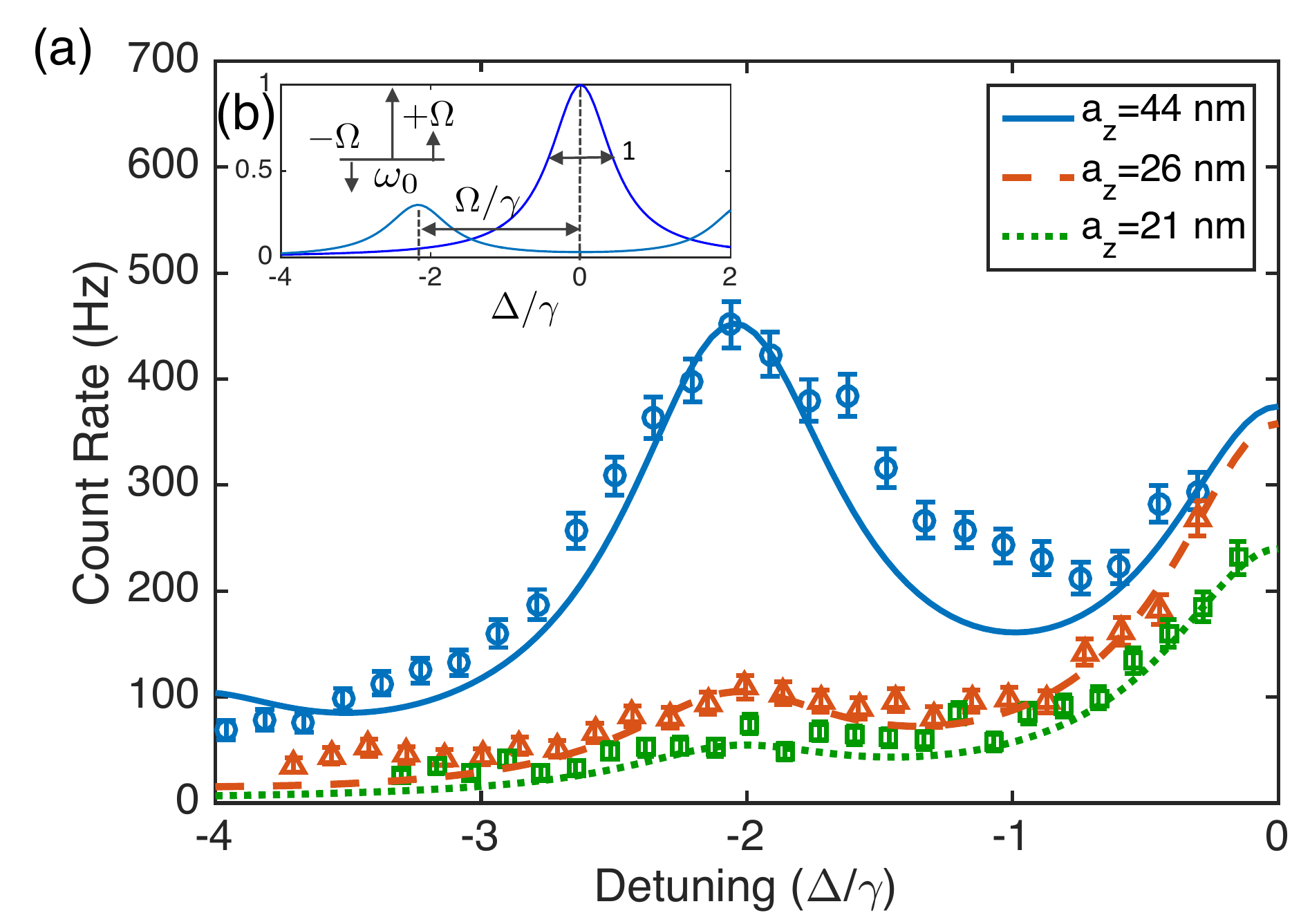}
	\caption{(Color online)  (a) The scattering rate of the ion as a function of laser detuning at the node of the SW for various levels of micromotion, where we use a fit to our model to estimate the micromotion amplitude.  (b) Schematic of the laser frequency in the ion's frame of reference, with micromotion sidebands at the RF frequency of $\pm \Omega$, and the expected frequency dependence of the ion's scattering rate. }
\label{fig:lineshapes} 
\end{figure}
Using the side fiber to collect the ion's fluorescence, we can probe the spatial field intensity $I(z)$ of the SW along the z-axis by monitoring the resonant scattering rate of the ion as its height above the trap is modulated by the tweaker electrode voltage.   Our model takes into account the ion's micromotion oscillations using the spatial probability density function of the ion over one period of its micromotion as $r(z')=1/\pi \sqrt{1/(a_{z}^2 - z'^2)}$ for $-a_z<z'<a_z$.  The ion's scattering rate in the SW can then be written as a convolution of the ion's position with the modulated scattering rate for an ion at rest as
\begin{equation}
\label{eqn:convolution}
S_{sw}(z,\Delta) =\int_{-a_z}^{a_z} r(z') S(z-z',\Delta) dz',
\end{equation}
where $S(z,\Delta)$ is the scattering rate from the modulated field intensity $I(z)$ with detuning $\Delta$.  The rate $S_{sw}(z,\Delta)$ is the quantity we measure with the side fiber coupling to determine the ion's relative height above the trap and is shown in Fig.~\ref{fig:oscillations}c as a function of the applied tweaker voltage for several attenuation levels.  Fig.~\ref{fig:oscillations}b shows the calibrated voltage amplitude on the tweaker electrodes for each level of DDS digital amplitude.  A direct measurement of the RF voltage applied to the tweaker electrodes contains a $\sim$10\% error due to stray capacitance, so we calibrate the RF voltage value on the trap by matching it with the trap simulation results.  Each data point in Fig.~\ref{fig:oscillations}c is the average count rate of fifty experiments, where the ion is exposed to the SW for 120 ms with 20 ms of Doppler cooling in between each measurement.  The RF trap position is moved symmetrically around the location of the DC null, which remains fixed during each scan.  The SW intensity was 40 mW/cm$^2$ and the beam was focused down to a waist radius of 5 $\mu$m on the mirror surface.  The minimum scattering rate at the node of the SW in Fig.~\ref{fig:oscillations}c did not significantly increase for an ion position within about one fringe on either side of the DC null, and is non-zero due to the presence of residual micromotion of the ion.  This is a result of charging of the dielectric mirror coatings upon exposure to UV light and is discussed below.

For the parameter regime we operate in where the trap frequencies are smaller than the spontaneous emission rate $\gamma$,  spontaneous emission must be taken into account as a damping force on the ion's motion.  To describe the scattering of the ion, we use a model of $S(z,\Delta)$ in Eqn.~\ref{eqn:convolution} which incorporates the modulation caused by the micromotion into the scattering rate in the low intensity limit \cite[]{PhysRevA.49.421}.  The stipulation for the low intensity limit of this model will hold near the node of the SW where the field approximately vanishes.  At the antinode, the field intensities can be on the order of $I_{sat}= 2\pi^2 \hbar \gamma c/3\lambda^3$ or greater and saturation effects cannot be ignored.  In general, these effects will cause a shift in the height, width, and frequency of the micromotion sidebands as the intensity is increased.\cite[]{micromotion_optical_clocks,PhysRevA.48.2169}  In the low intensity limit near the nodes where saturation effects are not expected, we use a simplified model for the scattering rate in terms of the carrier and micromotion sideband frequencies as
\begin{equation}
\label{eqn:scatN}
S(z, \Delta) \approx \frac{\gamma}{2} \frac{I(z)}{I_{sat}}  \sum_{n} \frac{J_n(\beta)^2}{1+(2(\Delta + n \Omega )/\gamma)^2}
\end{equation}
for ion heights $z$ such that $I(z) << I_{sat}$.  Substituting Eqn.~\ref{eqn:scatN} into Eqn.~\ref{eqn:convolution} gives a model that can be used to calculate $S_{sw}(z,\Delta)$ as a function of the ion's position and the laser detuning around the node of the SW.  Fig.~\ref{fig:lineshapes}a shows the measured SW scattering rate as a function of detuning at the node for several levels of mircomotion with amplitudes extracted from a fit of the data to our model.  Fig.~\ref{fig:lineshapes}b shows a schematic of the expected lineshape of the carrier and microwave sideband frequencies.  For the case of a large micromotion amplitude ($a_z \sim 40 $ nm), the lineshape begins to show some effects of saturation as the carrier and micromotion sideband transitions become wider and the sideband peak shifts in frequency relative to the predictions from our first order model.  Furthermore, we have found the micromotion sideband height at the node can increase above that of the anti-node for sufficiently large micromotion amplitudes.

\begin{figure}[t]
	\includegraphics[width=\columnwidth]{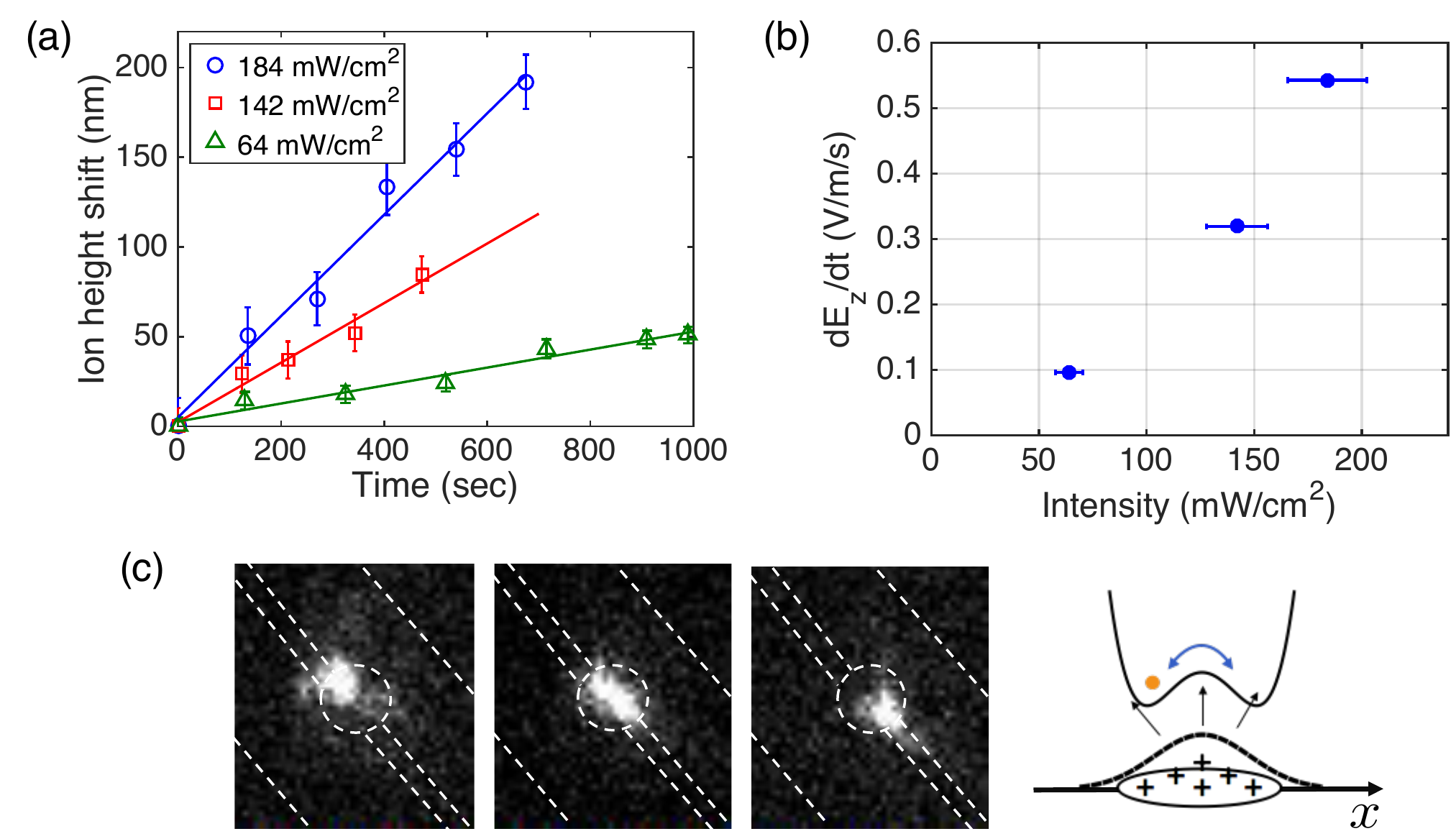}
	\caption{ (Color online) (a) The vertical displacement of the ion away from the RF null measured as a function of exposure time for various levels of SW beam intensity. (b) The rate of change of the electric field at the ion location as a function of the SW intensity. (c) Exposure to the SW for tens of minutes at a time can bifurcate the harmonic trapping potential into two minima along the axial trap direction.  Shown here are images of the ion located in each minima (first and third image) and directly in between (middle image) where it hops back and forth and smears the image.}
\label{fig:charging}
\end{figure}

The presence of micromotion which cannot be compensated in Fig.~\ref{fig:lineshapes}a is mainly attributed to charging of the dielectric mirror coatings.  Exposure of dielectric coatings to UV light, such as in the SW, is known to introduce positive charges into the material as electrons are ejected from the coating. \cite[]{charging_dielectrics}  This gives rise to a significant repulsive force directly beneath the ion, pushing it away from the area where the SW beam is incident on the mirror.  We model such a charge distribution with a two dimensional Gaussian function that is proportional to the incident laser intensity. 
We characterize the charging field at the ion as a function of the SW laser exposure time by measuring the displacement of the ion's position relative to the SW in response to the charging field, as shown in Fig.~\ref{fig:charging}a.  From the measured velocities of the three lines, we can find the rate of change of the electric field along the z-axis at the ion location as
\begin{equation}
\label{eqn:charge_rate}
\frac{d E_{z}(t)}{dt} = \frac{m \omega_z^2}{e} \frac{dz}{dt},
\end{equation}
where $\omega_z$ is the trap angular frequency along the z-axis and $m$ and $e$ are the mass and charge of the ion, respectively.  The rate of electric field change due to charging is shown in Fig.~\ref{fig:charging}b as a function of the SW intensity.  Due to the rapid increase of the charging rate as a function of the incident light intensity, the trap becomes unstable over long exposures to the point where the harmonic potential along the axial direction bifurcates into two potential minima as shown in Fig.~\ref{fig:charging}c.  

%While the optical intensity in the SW used here for precise determination of the ion's position is at least order of magnitude larger than what is required for current proposals to use the mirror trap with an optical cavity at UV wavelengths, the effects of charging must be taken into consideration the effects of charging of the mirror coating.  For photon collection experiments using ion fluorescence enhancement from an optical cavity, emitted photons are collected from the vacuum cavity mode (i.e. no external UV laser light is stored in the cavity), thereby significantly mitigating the effects of charging. 

In our current experiment, the SW serves as an accurate reference with which to measure the height of the ion above the trap.  In many important applications involving optical cavities, such as the ion fluorescence enhancement or efficient collection of photons emitted by the ion, no such strong UV light is expected in the cavity, and charging of the mirror coating will be largely mitigated.

In summary, we have demonstrated a key component of integrating a surface trap with an optical cavity by developing a trap made directly on a highly reflective UV mirror with the capability of adjusting the RF null position defining the trap location perpendicular to the trap surface with nanometer scale precision. 

This work was supported by ARO grant W911NF-15-1-0213.  Sandia National Laboratories is a multiprogram laboratory managed and operated by Sandia Corporation, a wholly owned subsidiary of Lockheed Martin Corporation, for the U.S. Department of Energy's National Nuclear Security Administration under Contract No. DE-AC04-94AL85000.

\bibliography{references.bib}{}

\end{document}